\documentclass[acmsmall,nonacm]{acmart}
\AtBeginDocument{%
  }

\usepackage[]{collab}
\usepackage{xcolor}
\usepackage{ulem}
\usepackage{pifont}
\usepackage{listings}
\usepackage{xcolor}
\usepackage{caption}
\usepackage{float}
\usepackage{hyperref}
\usepackage[]{collab}
\usepackage{balance}
\usepackage{xcolor}
\usepackage{subfigure}
\usepackage{multirow}
\usepackage{multicol}
\usepackage{tabularx}
\usepackage{siunitx}
\usepackage{float}
\usepackage{color}
\usepackage{diagbox}
\usepackage{caption}
\usepackage{makecell}
\usepackage{colortbl}
\usepackage{tcolorbox}
\usepackage{mdframed}
\usepackage{listings}
\usepackage{booktabs}
\usepackage[utf8]{inputenc}
\usepackage{calligra}
\usepackage[normalem]{ulem}
\usepackage{url}
\usepackage{soul}
\usepackage{nicematrix}
\usepackage{fontawesome5}
\usepackage{textcomp}
\usepackage{stfloats}
\usepackage{verbatim}
\usepackage{graphicx}
\usepackage{amsmath,amsfonts}
\usepackage{algorithmic}
\usepackage{array}
\usepackage[utf8]{inputenc}
\usepackage{enumitem}

\collabAuthor{js}{red}{Junsong}
\collabAuthor{zb}{teal}{Zhuangbin}

\newcommand\abcoder{\textsc{ABCoder}\xspace}

\definecolor{tsKeyword}{RGB}{0,77,140}
\definecolor{tsComment}{RGB}{92,99,112}
\definecolor{tsString}{RGB}{163,87,27}

\lstdefinelanguage{TypeScript}{
  morekeywords={
    async,await,class,const,export,from,function,import,new,return
  },
  morecomment=[l]{//},
  morestring=[b]",
  morestring=[b]',
  sensitive=true
}

\lstdefinestyle{tscode}{
  basicstyle=\ttfamily\small,
  frame=single,
  breaklines=true,
  commentstyle=\color{tsComment},
  keywordstyle=\color{tsKeyword}\bfseries,
  numbers=left,
  numbersep=8pt,
  numberstyle=\scriptsize\color{tsComment},
  showstringspaces=false,
  stringstyle=\color{tsString}
}

\setcopyright{none}

\begin{document}

\title{TypeScript Repository Indexing for Code Agent Retrieval}

\author{Junsong Pu}
\affiliation{%
  \institution{Sun Yat-sen University}
  \city{Zhuhai}
  \country{China}}
\email{pujs@mail2.sysu.edu.cn}

\author{Yichen Li}
\affiliation{%
  \institution{The Chinese University of Hong Kong}
  \city{Hong Kong}
  \country{China}}
\email{ycli21@cse.cuhk.edu.hk}

\author{Zhuangbin Chen}
\authornote{Zhuangbin Chen is the corresponding author.}
\affiliation{%
  \institution{Sun Yat-sen University}
  \city{Zhuhai}
  \country{China}
  }
\email{chenzhb36@mail.sysu.edu.cn}

\renewcommand{\shortauthors}{Pu et al.}

\begin{abstract}
Graph-based code indexing can improve context retrieval for LLM-based code agents by preserving call chains and dependency relationships that keyword search and similarity retrieval often miss. \abcoder is an open-source framework that parses codebases into a function-level code index called UniAST. Its existing parsers combine lightweight AST parsers for syntactic analysis with language servers for semantic resolution, 
but because LSP-based resolution requires a JSON-RPC call for each symbol lookup, these per-symbol calls become a bottleneck on large TypeScript repositories. 
We present \texttt{abcoder-ts-parser}, a TypeScript parser built on the TypeScript Compiler API that works directly with the compiler's AST, semantic information, and module resolution logic. We evaluate the parser on three open-source TypeScript projects with up to 1.2 million lines of code and find that it produces reliable indexes significantly more efficiently than the existing architecture. For a live demonstration, watch: \url{https://youtu.be/ryssr7ouvdE}
\end{abstract}




\maketitle

\section{Introduction and Background}

\subsection{Context Retrieval for Code Agent}

\subsubsection{Code Agent}
Recent advances in Large Language Model (LLM) have given rise to a new class of development tools known as \textit{LLM-based code agents}. Systems such as Claude Code~\cite{AnthropicClaudeCode2026} operate as autonomous assistants capable of performing a wide range of software engineering tasks with minimal human intervention. At their core, these agents follow a ReAct-style~\cite{Yao2022ReAct} loop of reasoning and acting: given a developer's request, the agent first gathers relevant context from the project, then reasons over the gathered information to decide on an action, observes the outcome, and loops back if the result is unsatisfactory. This cycle repeats until the task is completed or the agent determines that human input is needed.

\subsubsection{Efficient Context Retrieval}
The effectiveness of a code agent depends heavily on the quality of its \textit{context retrieval}, which we define as the process of locating and assembling the relevant code fragments that the model needs to reason about a given task. Benchmarks and systems on real repository-level software engineering tasks consistently show that solving a single issue often requires coordinating changes across multiple functions, classes, and files, while identifying only a subset of repository context as truly relevant~\cite{errorprism, logimprover,Yang2024SWEAgent,Xia2024Agentless,Ouyang2024RepoGraph}. Because a language model can only reason about the code that fits within its prompt, poor retrieval leads to incomplete or misleading input, which in turn degrades the model's reasoning even when its underlying capabilities are sufficient~\cite{Ouyang2024RepoGraph}. For example, when fixing a bug that spans multiple functions across several files, the model needs to see not only the function where the error manifests but also the functions that call it and the types it depends on;
if the retrieval step returns only the direct function, the model often fails to find the root cause and might suggest an incomplete or incorrect patch. 
In this sense, the quality of context retrieval directly shapes the quality of the model's downstream reasoning, and improving retrieval is a practical path to improving agent performance on complex software engineering tasks.

Existing approaches to context retrieval in code agents fall broadly into three categories. The first is \textit{command-line exploration}, where the agent executes shell commands such as \texttt{grep}, \texttt{find}, or other utilities to locate relevant files and symbols. 
Software engineering agents like SWE-agent~\cite{Yang2024SWEAgent} often rely on this method to navigate codebases.
It has the notable advantage of requiring no upfront indexing: it works on any codebase immediately without building or maintaining any auxiliary data structures, making it the simplest strategy to deploy. However, command-line exploration is fundamentally keyword-driven, which causes problems in both directions: it misses relevant code that does not share lexical overlap with the query, and it returns irrelevant results that happen to contain the same keywords, forcing the agent to spend additional rounds filtering out noise. It also places a heavy burden on the agent's planning ability, as the agent needs to decide which commands to run, interpret their outputs, and iteratively refine its search, often consuming many interaction rounds before arriving at useful results. This burden has motivated alternative pipelines such as Agentless~\cite{Xia2024Agentless}, which separate localization from repair rather than relying on open-ended interaction throughout the entire loop.

The second is similarity-based retrieval, where systems fetch relevant code snippets using either keyword matching algorithms or semantic vector embeddings.
Similarity-based retrieval-augmented systems such as RepoCoder~\cite{Zhang2023RepoCoder} and Repoformer~\cite{Wu2024Repoformer} show that this strategy can improve repository-level completion quality over purely local baselines. 
While similarity retrieval is effective at matching natural language queries to relevant code snippets, it is less suited to reasoning about function behavior, which requires capturing structural relationships such as call chains and type dependencies that similarity representations do not encode.
Large-scale analyses such as CodeRAG-Bench~\cite{Wang2025CodeRAGBench} further show that current similarity retrievers still struggle to fetch useful contexts when lexical overlap is weak, and downstream models still need to further analyze the retrieved fragments to reason about the relationships between code elements.

The third category is graph-based code indexing, where a parser performs static analysis on a repository to extract entities such as functions, classes, and modules along with their call and dependency relationships, and organizes them into a queryable graph. This approach preserves the structural information that the previous two methods discard. Recent graph-based retrieval systems such as RepoGraph~\cite{Ouyang2024RepoGraph} explicitly model repository structure for downstream software engineering agents. \abcoder~\cite{CloudWeGoABCoder2026} is a representative system in this category: it parses a codebase into a function-level graph and builds a Graph RAG that allows an agent to retrieve not just a single function but also its related code entities, including callers, callees, and related type definitions. 

\subsection{\abcoder Framework}
\label{subsec:abcoder}
\subsubsection{Overview}
\abcoder is a toolset that builds graph-based code index from source code and provides query utilities to improve context retrieval for code agents. Given a repository, it parses the code into a language-agnostic format called UniAST (Universal Abstract Syntax Tree). This format captures code entities like functions, types, and variables alongside their dependencies. LLM agents can then use \abcoder's query tools via the Model Context Protocol (MCP) ~\cite{AnthropicMCP2024} to navigate the resulting code index for tasks like codebase exploration and cross-file dependency tracing. By supplying a relational graph instead of raw text, the toolset helps the model analyze code as a network of connected entities rather than isolated snippets.

\subsubsection{Performance on Large Repositories}
Existing \abcoder parsers combine the Language Server Protocol (LSP)~\cite{MicrosoftLSP2026} with lightweight AST parsers such as Tree-sitter~\cite{TreeSitter2026}. In this architecture, Tree-sitter scans each source file to identify all code entities and potential dependency sites such as call expressions, type references, and import statements. Since Tree-sitter is a syntax-level parser and does not resolve symbols to their definitions, it cannot determine what a given symbol actually refers to. For each dependency site, the parser therefore sends a request to the language server, which communicates via \texttt{JSON-RPC}, to resolve the symbol's definition. The language server returns a file location, which the parser maps back to an entity in Tree-sitter's syntax tree of the target file, thereby establishing an edge in the dependency graph. 

This architecture borrows from how most modern code editors work: Tree-sitter provides fast syntax highlighting and structural navigation, while the language server handles semantic queries on demand. It is also largely language-agnostic, as adding support for a new language mainly requires installing the corresponding language server and Tree-sitter grammar, without changing the indexing framework itself. When a developer hovers over a symbol in an editor to see its definition, the cost of resolving one symbol at a time is negligible.

\begin{lstlisting}[caption={A minimal example for TypeScript indexing.},label={lst:ts-indexing-example},language=TypeScript,style=tscode]
// user-repo.ts
export class UserRepo {
  getById(id: string) { return id; }
}

// index.ts
export { UserRepo as Repo } from "./user-repo";

// service.ts
import { Repo } from "./index";
export function loadUser(id: string) {
  const repo = new Repo();
  return repo.getById(id);
}
\end{lstlisting}

For large repository indexing, however, this cost accumulates quickly. Listing~\ref{lst:ts-indexing-example} illustrates this with a small TypeScript example. Tree-sitter scans \texttt{service.ts} and identifies the function \texttt{loadUser} as an entity, along with three dependency sites: the import of \texttt{Repo}, the constructor call \texttt{new Repo()}, and the method call \texttt{repo.getById(id)}. 
Each of these sites requires at least one language-server request to resolve, and the import of \texttt{Repo} requires an additional request, because it first resolves to the \textit{barrel file} \texttt{index.ts}, a module that re-exports symbols from other files to provide a single entry point for the package, and a follow-up request is needed to trace the re-export to \texttt{UserRepo} in \texttt{user-repo.ts}.
Each response must also be mapped back to Tree-sitter's syntax tree again. For a large TypeScript project with thousands of functions, the total number of RPC calls can scale significantly. Since the LSP protocol resolves symbols one position at a time and does not offer batch resolution, this makes repository indexing slow for large TypeScript projects. To avoid this, we build our parser directly on the TypeScript Compiler API, which loads the entire project into a single in-process compiler instance and provides the AST, semantic information, and module resolution results without any external RPC calls.

\section{System Design}

We submitted the first version of \texttt{abcoder-ts-parser} to the \abcoder project in September 2025. Since then, we have been maintaining and developing the tool in collaboration with the \abcoder open-source community. As of April 2026, the parser contains about 6 KLOC of TypeScript code, excluding test code. This section describes its usage and output format, followed by how the parser reduces the indexing overhead discussed earlier.

\subsection{Tool Overview}

\subsubsection{Usage} Our TypeScript parser is distributed as a standalone command-line tool called \texttt{abcoder-ts-parser}. Given a TypeScript project directory and the paths to its metadata files, it performs a one-pass static analysis and produces a single JSON file conforming to \abcoder's UniAST specification~\cite{CloudWeGoUniAST2026}. The tool also supports a monorepo mode for repositories that contain multiple packages under a single root. During parsing, the tool reads these metadata files to determine project configuration and dependencies.

\subsubsection{Output Format} The output follows UniAST's hierarchical structure, summarized in Table~\ref{tab:uniast-structure}. At the top level, the output contains a set of modules and a repository code index. In a monorepo, each constituent project is represented as a separate module; in a single-project repository, there is exactly one module that corresponds to the repository itself. Each module contains one or more packages, where each package maps to a single source file. Each package contains three kinds of entities: functions, types, and variables. For every entity, the parser records its unique identity (a triple of module path, package path, and symbol name), its complete source text, its file location, and its signature.

\begin{table}[h]
\centering
\small
\caption{UniAST output structure.}
\label{tab:uniast-structure}
\begin{tabular}{lp{5.2cm}}
\hline
\textbf{Level} & \textbf{Contents} \\
\hline
\textit{Repository} & Modules, repository code index \\
\hline
\textit{Module} & Packages, Dependencies, Files \\
\hline
\textit{Package} & Functions, Types, Variables \\
\hline
\textit{Entity} & Identity, source text, file location, signature, per-entity dependency list \\
\hline
\end{tabular}
\end{table}

The repository code index is what distinguishes UniAST from a flat collection of code snippets. It aggregates the dependency lists of all entities into a global structure. Each node in this structure represents one code entity, and each edge encodes a relationship between two entities. Table~\ref{tab:relations} describes the four kinds of relations captured by the graph.

\begin{table}[h]
\centering
\small
\caption{Relation types in the UniAST dependency graph.}
\label{tab:relations}
\begin{tabular}{lp{5.2cm}}
\hline
\textbf{Relation} & \textbf{Semantics} \\
\hline
\textit{Dependency} & Entity \textit{A} calls, references, or otherwise depends on entity \textit{B}. \\
\hline
\textit{Reference} & The inverse of Dependency: entity \textit{B} is depended upon by entity \textit{A}. \\
\hline
\textit{Implementation} & A class or object implements the contract defined by an interface. \\
\hline
\textit{Group} & Entities declared together as a logical unit, e.g., members of a single \texttt{enum} or \texttt{const} block. \\
\hline
\end{tabular}
\end{table}

These relations support different forms of structured retrieval. Dependency and Reference edges form a bidirectional link between related entities, allowing a code agent equipped with a code index retrieval tool to traverse the graph in both directions. For example, given a function that a developer is asking about, the agent can follow its Dependency edges outward to gather the functions it calls and the types it uses, and follow its Reference edges inward to gather the functions that call it. This directly provides the agent with the relevant call and dependency information for that entity, rather than returning isolated fragments that happen to share similar keywords or embeddings. Implementation edges connect interface definitions to their concrete implementations, which is useful in TypeScript codebases where components are often consumed through abstract interfaces.

\subsection{Parsing with the TypeScript Compiler API}

As discussed in Section~\ref{subsec:abcoder}, resolving symbols and dependencies through a separate language server requires many RPC calls between the parser and the server, one for each dependency site in each file. Our parser avoids this by loading the entire project into a single compiler instance, from which the AST, semantic information, and module resolution results are all directly available in memory. 
The parser walks the compiler's AST and reduces it into a code index following the UniAST specification. It identifies functions, types, and variables, records their signatures, source text, and locations, and extracts the dependency relationships between them, including direct calls, method calls, constructor invocations, and type references. 
During this process, a symbol resolver traces each referenced symbol along its import chain, following aliases, re-exports, and barrel files until it reaches the original declaration, so that every dependency edge in the output graph points to the actual definition rather than an intermediate re-export.

\section{Evaluation}

\subsection{Testing and Validation}

For a code parser, validation focuses on whether entities are identified accurately, dependency edges point to the intended targets, and cross-file symbol resolution follows import chains to the actual definition. Unit testing is a natural fit for checking these properties, as each parsing rule can be tested against specific code patterns with known expected outputs. We therefore use the parser's unit test suite and its code coverage metrics as our primary validation evidence.

Since its initial release in September 2025, \texttt{abcoder-ts-parser} has been under continuous development by the \abcoder open-source community. Over the course of eight months, the test suite has grown to 10 test suites comprising 175 individual test cases and about 7 KLOC of test code, of which 174 currently pass. The single failing test is due to a discrepancy in how monorepo root packages are counted and does not affect the core parsing logic.

\begin{table}[h]
\centering
\small
\caption{Test coverage of key modules.}
\label{tab:coverage}
\begin{tabular}{lrr}
\hline
\textbf{Module} & \textbf{Line Cov.} & \textbf{Branch Cov.} \\
\hline
FunctionParser & 85.36\% & 52.00\% \\
\hline
TypeParser & 93.18\% & 76.10\% \\
\hline
VarParser & 68.66\% & 50.92\% \\
\hline
ModuleParser & 87.05\% & 58.06\% \\
\hline
PackageParser & 100\% & 100\% \\
\hline
graph-builder & 98.75\% & 95.65\% \\
\hline
symbol-resolver & 77.08\% & 66.40\% \\
\hline
\textit{Overall} & \textit{71.75\%} & \textit{51.73\%} \\
\hline
\end{tabular}
\end{table}

Table~\ref{tab:coverage} reports the line and branch coverage for the major components of the parser. The core parsing and graph construction modules generally achieve line coverage above 80\%, with several modules reaching above 93\%. The overall line coverage across all source files is 71.75\%, with the lower figure mainly due to infrastructure code such as the CLI entry point and the parallel processing layer, which are not easily covered by unit tests alone.

\subsection{Parsing Performance}

To evaluate the parser's efficiency on real-world codebases, we benchmark it on three open-source TypeScript projects of varying scale: Excalidraw~\cite{ExcalidrawRepo2026}, an online collaborative whiteboard application; Outline~\cite{OutlineRepo2026}, a team knowledge base and document collaboration platform; and Sentry~\cite{SentryRepo2026}, a large-scale error monitoring and performance tracking platform. All experiments run on a machine with an Intel Xeon Platinum 8336C CPU (32 cores at 2.30\,GHz) and 62\,GB of RAM, with the Node.js process allocated a 16\,GB heap.

\begin{table}[h]
\centering
\small
\caption{Parsing performance on three open-source TypeScript projects.}
\label{tab:performance}
\begin{tabular}{lrrr}
\hline
\textbf{Repository} & \textbf{Files} & \textbf{Lines of Code} & \textbf{Time (s)} \\
\hline
Excalidraw & 601 & 147K & 35.2 \\
\hline
Outline & 1{,}875 & 234K & 239.9 \\
\hline
Sentry & 8{,}909 & 1.23M & 705.6 \\
\hline
\end{tabular}
\end{table}

As shown in Table~\ref{tab:performance}, the parser handles projects ranging from several hundred to nearly nine thousand source files. For a large project like Sentry, the largest project at over 1.2 million lines of code, the full analysis completes in under 12 minutes. Since the index only needs to be rebuilt when the codebase changes significantly, this one-time cost is acceptable in practice. For smaller projects like Excalidraw, the parsing finishes in about 35 seconds, making the tool suitable for integration into regular development workflows. As a point of comparison, \abcoder's existing Python parser, which uses a lightweight AST parser plus LSP approach, takes 603 seconds to parse Django (160 KLOC) under the same hardware setting. Our parser analyzes a comparable amount of TypeScript code (Excalidraw, 147KLOC) in 35 seconds, suggesting that avoiding individual symbol-resolution RPC calls yields a substantial reduction in parsing time.

\subsection{Downstream Usage}

Once the parser produces a UniAST index, it can serve a variety of downstream purposes. The primary use case is graph-based retrieval for code agents. \abcoder provides an MCP server that loads the index and exposes a set of query tools, allowing agents such as Claude Code to navigate the codebase and retrieve a specific function along with its callers, callees, and type dependencies. Because the index is organized at function-level granularity, each retrieval step returns a semantically meaningful unit of code rather than an arbitrary text chunk, reducing irrelevant context in the model's prompt.
Another use case is automated codebase documentation. The index provides a precomputed context slice for each function, containing its callers, callees, and type dependencies, so the language model can describe the function's role directly without scanning raw source files to discover these relationships.

\section{Data Availability}

\texttt{abcoder-ts-parser} is publicly available in the \abcoder repository on GitHub at \url{https://github.com/cloudwego/abcoder/tree/main/ts-parser}~\cite{CloudWeGoABCoder2026}. Also available on npm: \url{https://www.npmjs.com/package/abcoder-ts-parser}

\begin{acks}
   We thank the CloudWeGo team for their support, especially Wenju Gao, Zhenyang Dai, Xuran Yin, Zekun Wang, and Yi Duan for their sustained collaboration.
\end{acks}

\bibliographystyle{ACM-Reference-Format}
\bibliography{sample-base}

\end{document}